\documentclass[aps,prb,a4paper,10pt,twocolumn,showpacs,floatfix,superscriptaddress,preprintnumbers,longbibliography,bibnotes]{revtex4-2}
\usepackage{float}
\usepackage{booktabs}

\usepackage{dcolumn,graphicx,color,booktabs,microtype,afterpage}
\usepackage{amssymb}
\usepackage{amsmath}
\usepackage[charter,greekuppercase=italicized]{mathdesign}
\usepackage{sidecap}
\usepackage[mathlines]{lineno}

\graphicspath{{./}{figure/}}
\renewcommand{\tablename}{Table}
\makeatletter\renewcommand{\fnum@figure}[1]{\figurename~\thefigure.~}\makeatother
\makeatletter\renewcommand{\fnum@table}[1]{\tablename~\thetable.}\makeatother

\newcount\hh \newcount\mm
\hh=\time \divide\hh by 60
\mm=\hh \multiply\mm by 60 \mm=-\mm
\advance\mm by \time
\def\now{\number\hh:\ifnum\mm<10{}0\fi\number\mm}

\usepackage[colorlinks,plainpages=false,linkcolor=blue,urlcolor=blue,citecolor=blue,pdfpagemode=UseNone,pdfstartview=FitBH]{hyperref}
\usepackage{nicefrac}

\begin{document}

\makeatletter\renewcommand{\ps@plain}{%
\def\@evenhead{\hfill\itshape\rightmark}%
\def\@oddhead{\itshape\leftmark\hfill}%
\renewcommand{\@evenfoot}{\hfill\small{--~\thepage~--}\hfill}%
\renewcommand{\@oddfoot}{\hfill\small{--~\thepage~--}\hfill}%
}\makeatother\pagestyle{plain}

\title{Nodeless superconductivity in the noncentrosymmetric ThIrSi compound}

\author{D.~Tay}
\affiliation{Laboratorium f\"ur Festk\"orperphysik, ETH Z\"urich, CH-8093 Z\"urich, Switzerland}

\author{T.~Shang}\email[Corresponding authors:\\]{tshang@phy.ecnu.edu.cn}

\affiliation{Key Laboratory of Polar Materials and Devices (MOE), School of Physics and Electronic Science, East China Normal University, Shanghai 200241, China}
%

\author{Priscila~F.~S.~Rosa}
\affiliation{Los Alamos National Laboratory, Los Alamos, NM 87545, USA}
\author{F.~B.~Santos}
\affiliation{Los Alamos National Laboratory, Los Alamos, NM 87545, USA}
\affiliation{Escola de Engenharia de Lorena, Universidade de S\~{a}o Paulo (EEL-USP), Materials Engineering Department (Demar), Lorena, S\~ao Paulo, Brazil}
%
%
%
\author{J.~D.~Thompson}
\affiliation{Los Alamos National Laboratory, Los Alamos, NM 87545, USA}
\author{Z. Fisk}
\affiliation{Department of Physics and Astronomy, University of California at Irvine, Irvine, California 92697, USA}
\author{H.-R.~Ott}
\affiliation{Laboratorium f\"ur Festk\"orperphysik, ETH Z\"urich, CH-8093 Z\"urich, Switzerland}
\author{T.~Shiroka}\email{tshiroka@phys.ethz.ch}
\affiliation{Laboratorium f\"ur Festk\"orperphysik, ETH Z\"urich, CH-8093 Z\"urich, Switzerland}
\affiliation{Laboratory for Muon-Spin Spectroscopy, Paul Scherrer Institut, Villigen PSI, Switzerland}
\begin{abstract}
The ThIrSi superconductor, with $T_c = 6.5$\,K, is expected to show
unusual features in view of its noncentrosymmetric structure and the
presence of heavy elements featuring a sizable spin-orbit coupling.
Here, we report a comprehensive study of its electronic properties
by means of local-probe techniques: muon-spin rotation and relaxation
({\textmu}SR) and nuclear magnetic resonance (NMR). Both the superfluid
density $\rho_\mathrm{sc}(T)$ (determined via transverse-field {\textmu}SR)
and the spin-lattice relaxation rate $T_1^{-1}(T)$ (determined via NMR)
suggest a nodeless superconductivity. Furthermore, the absence of
spontaneous magnetic fields below $T_c$, as evinced from zero-field
{\textmu}SR measurements, indicates a preserved time-reversal symmetry in
the superconducting state of ThIrSi. Temperature-dependent upper critical
fields as well as field-dependent superconducting muon-spin relaxations
suggest the presence of multiple superconducting gaps in ThIrSi.
\end{abstract}

\maketitle\enlargethispage{3pt}

\vspace{-5pt}
\section{Introduction}\enlargethispage{8pt}
\label{sec:Introduction} 

Superconductors whose crystal structures lack an inversion center are
known as noncentrosymmetric superconductors (NCSCs) and represent
appealing systems for investigating unconventional- and topological
superconductivity (SC)~\cite{Bauer2012,Smidman2017,Ghosh2020b,Kim2018,Sun2015,Ali2014,Sato2009,Tanaka2010,Sato2017,Qi2011,Kallin2016}.
Since in NCSCs parity is not a good quantum number, this allows for the
presence of an antisymmetric spin-orbit coupling (ASOC), which lifts the
degeneracy of the conduction-electron bands and splits the Fermi surface.
Consequently, both intra- and inter-band Cooper pairs can be formed,
resulting in an admixture of spin-singlet and spin-triplet
pairings~\cite{Bauer2012,Smidman2017}. Unfortunately, in actual NCSCs
systems, while some of them, like Li$_{2}$Pt$_{3}$B, do indeed exhibit
spin-triplet pairing \cite{Shimamura2007nmr}, many others do not. For
instance, previous studies on some highly-anticipated NCSCs, such as
Re$_7$B$_3$\cite{Shang2021multigap}, W$_3$Al$_2$C~\cite{Tay2022},
Mo$_3$Al$_2$C\cite{Bauer2014} and NbReSi~\cite{Shang2022NbReSi}, reveal
no spin-triplet superconductivity, despite sizable SOC interactions.

However, spin-triplet superconductivity is not the only possible outcome
of ASOC. Since ASOC causes the splitting of the Fermi surface, it may
conceivably also constitute a generic mechanism to achieve two-band
superconductivity. The latter has been a prominent issue following the
discovery of superconductivity in MgB$_2$~\cite{Akimitsu2001}, where the
presence of strongly anisotropic $\sigma$-bands and isotropic $\pi$-bands 
gives rise to two-band SC~\cite{Souma2003origin}. Since then, efforts
have been made to investigate and understand two-band superconductivity
also in other materials. Among the most promising two-band superconductors
are the sesquicarbides, which comprise the La$_2$C$_3$ and Y$_2$C$_3$
NCSCs. NMR \cite{Harada2007} and {\textmu}SR studies~\cite{Kuroiwa2008} of
sesquicarbides provide hints of multigap superconductivity, with both
bands proposed to be of $s$-type, thus implying a two-band $(s+s)$ model.
On the other hand, heat-capacity-\cite{Akutagawa2007} and tunneling
break-junction measurements~\cite{Ekino2013} on the same compounds
indicate a conventional $s$-type single-band superconductivity. Hence,
the question of whether the sesquicarbides can be described by a
two-band- or by a single-band model is still controversial. Such
controversy is largely due to the fact that, in the proposed two-band
model of sesquicarbides, the average gap value is close to the BCS
theoretical value $2\Delta/k_\mathrm{B}T_c = 3.52$, making it
experimentally challenging to distinguish between the two cases.
A two-band model was also used to describe the {\textmu}SR results in 
NbReSi~\cite{Shang2022NbReSi}, where again the two bands have similar
gap values, so the additional band does not significantly affect the
physical properties of this system. 
As for ThIrSi, object of the current study, previous DFT calculations
have shown the presence of multiple bands near the Fermi
surface~\cite{ptok2019electronic}, hence strongly suggesting that also
ThIrSi may exhibit multiband superconductivity. 

Ternary equiatomic transition-metal silicides of the Th$T\!$Si
family, with $T$ = Co, Ir, Ni, and Pt, have been known since the
eighties \cite{Klepp1982,SubbaRao1985,Zhong1985,Chevalier1986}. The
original focus was on their synthesis and structural characterization,
followed by the detection of superconductivity through electrical resistivity
measurements~\cite{SubbaRao1985,Zhong1985}. Later on, these first results
were complemented by more detailed specific-heat and contact-point
spectroscopy investigations~\cite{Chevalier1986}. In recent years, interest
in ternary transition-metal silicides was re-ignited by the discovery
of spin-triplet SC in La$T$(Si,Ge) Weyl nodal-line semimetals, whose
time-reversal symmetry (TRS) is broken in the superconducting
state~\cite{Shang2022spin}. Yet, less is known about their thorium
conterpart.

Here, we revisit the Th$T\!$Si family in more detail and report about
a comprehensive study of superconductivity in ThIrSi by means of SQUID
magnetometry, {\textmu}SR, and NMR. Below we show that ThIrSi is an
unambiguous example of a NCSC that exhibits nodeless superconductivity.
All our results show features typical of fully-gapped superconductors.
The temperature-dependent superfluid density is compatible with either
a single-band $s$-wave model with $\Delta/k_\mathrm{B}T_c =  2.10(5)$,
or with a two-band $(s+s)$ model, with
$\Delta_{\alpha} = 1.90(5) k_\mathrm{B} T_{c}$ and
$\Delta_{\beta} = 2.20(5) k_\mathrm{B} T_{c}$.
This uncertainty about multiband nature is resolved by
measurements of temperature-dependent upper critical fields and of
field-dependent muon-spin relaxation in the SC state, both of
which clearly suggest multiband superconductivity in ThIrSi.

\section{Experimental details}\enlargethispage{8pt}
\label{sec:details}

\begin{figure}
	\centering
	\includegraphics[width=0.47\textwidth,angle=0]{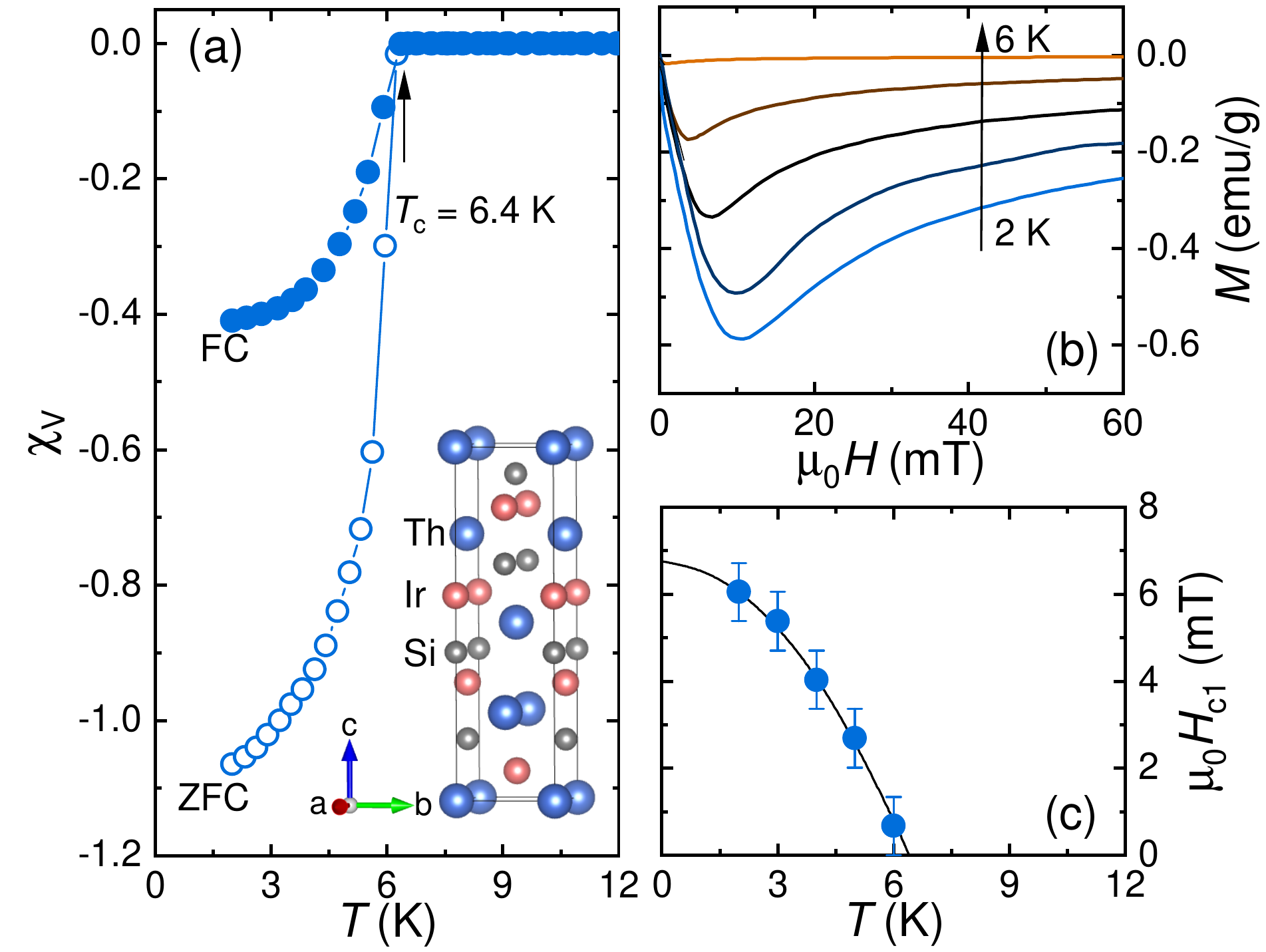}
	\vspace{-2ex}%
	\caption{\label{fig:chi}(a) Temperature dependence of the magnetic
	susceptibility $\chi_\mathrm{V}(T)$, in SI units, measured at 1\,mT.
	The inset shows the crystal structure of ThIrSi.
	(b) Field-dependent magnetization curves collected at various
	temperatures after cooling the sample in zero field.
	(c) Lower critical fields $H_\mathrm{c1}$ vs.\ temperature. Solid
	lines are fits to $\mu_{0}H_\mathrm{c1}(T) =\mu_{0}H_\mathrm{c1}(0)[1-(T/T_{c})^2]$.
	For each temperature, $H_\mathrm{c1}$ was determined as the value
	where $M(H)$ starts deviating from linearity. Both magnetic
	susceptibility values and lower critical fields were corrected by
	accounting for the demagnetization factor.}
\end{figure}

Polycrystalline ThIrSi samples were prepared by arc melting 
stoichiometric amounts of Th (99.9\%), Ir (99.95\%), and Si (99.9999\%)
in a water-cooled copper hearth under argon atmosphere. No weight loss
was observed during the melting process. The obtained arc-melted button
was flipped over and melted repeatedly to ensure homogeneity. The as-cast
samples were then wrapped in a tantalum foil and annealed in a quartz
tube under vacuum at 1000$^\circ$C for one week. The crystal structure
of the resulting alloy was checked at ambient temperature by means of
powder x-ray diffraction using Cu K$\alpha$ radiation. This confirmed a
noncentrosymmetric tetragonal structure of LaPtSi-type with space group
$I4_1md$ (No.~109) [see inset in Fig.~\ref{fig:chi}(a)].
Magnetization measurements were performed on a Quantum Design magnetic
property measurement system.

The bulk {\textmu}SR measurements were carried out at the ge\-ne\-ral\--pur\-pose
surface-muon instrument (GPS) of the Swiss muon source at Paul Scherrer
Institut, Villigen, Switzerland. 
In this study, we performed two types of experiments: transverse-field 
(TF)-, and zero-field (ZF)-{\textmu}SR measurements. 
As to the former, it allowed us to determine the temperature evolution
of the superfluid density. As to the latter, we aimed at searching
for a possible breaking of time-reversal symmetry in the superconducting
state of ThIrSi. 
To avoid the effects of stray magnetic fields during the ZF-{\textmu}SR
measurements, all magnets were preliminarily degaussed. The {\textmu}SR
spectra were collected upon sample heating and then analyzed by means
of the \texttt{musrfit} software package~\cite{Suter2012}.

The $^{29}$Si NMR measurements, including line shapes and spin-lattice 
relaxation times, were performed on ThIrSi in powder form in a magnetic
field of 3\,T. The NMR reference frequency $\nu_0$ was determined from
the $^{29}$Si resonance signal in tetramethylsilane (TMS). 
Subsequently, the $^{29}$Si NMR shifts were calculated 
with respect to $\nu_0$. 
%
%
To cover the 2 to 300\,K temperature range we used a continuous-flow 
CF-1200 cryostat by Oxford Instruments, with temperatures below 4.2\,K 
being achieved under pumped $^{4}$He conditions. 
The $^{29}$Si NMR signal was detected by means of a standard spin-echo 
sequence consisting of $\pi/2$ and $\pi$ pulses of 3 and 6\,{\textmu}s, 
with recycling delays ranging from 1 to 60\,s in the 2--300\,K
temperature range. 
The line shapes were obtained via fast Fourier transform (FFT) of 
the echo signal. Spin-lattice relaxation times $T_1$ were measured 
via the inversion-recovery method, using a $\pi$--$\pi/2$--$\pi$ 
pulse sequence. In all the measurements, phase cycling was used to 
systematically minimize the presence of artifacts.

\begin{figure}[!htp]
	\centering
	\includegraphics[width=0.49\textwidth,angle= 0]{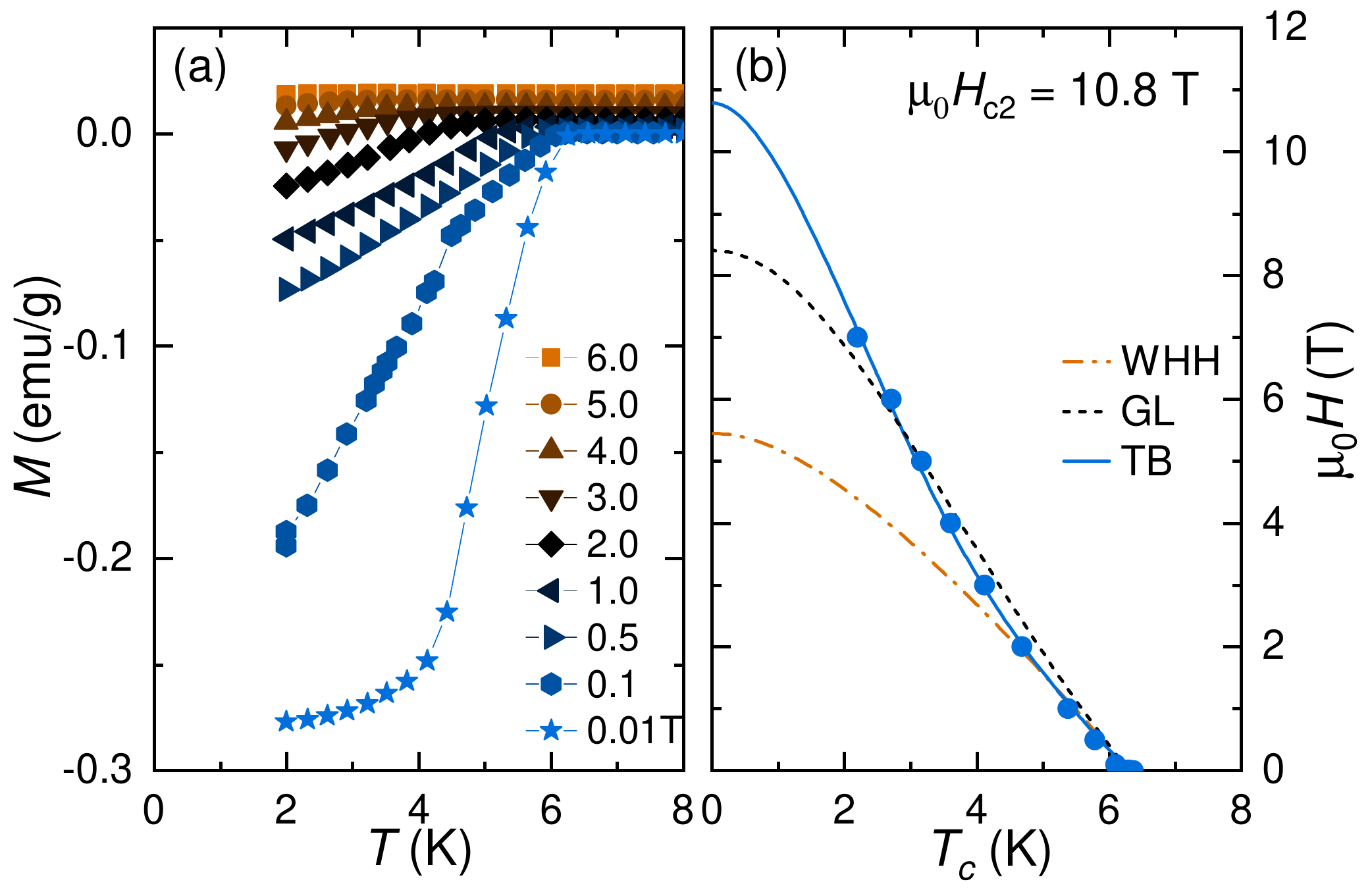}
	\caption{\label{fig:Hc2}(a) Temperature-dependent magnetization
	curves $M(T,H)$ for various applied magnetic fields. (b) Upper
	critical fields $H_\mathrm{c2}$ vs.\ transition temperature $T_c$. 
    The dash-dotted-, dotted-, and solid lines represent fits to the
    WHH, GL, and TB models, respectively.}
\end{figure}

\section{Results and discussion}\enlargethispage{8pt}
\label{sec:results}

\subsection{Magnetization measurements}
The bulk superconductivity of ThIrSi was first characterized by
magnetic-susceptibility measurements,
using both field-cooled (FC) and zero-field-cooled (ZFC) protocols
in an applied field of 1\,mT. As shown in Fig.~\ref{fig:chi}(a),
a clear diamagnetic signal appears below the superconducting transition
at $T_c$ = 6.4\,K. A rather sharp transition (with a $\Delta T \sim 0.4$\,K)
indicates a good sample quality. After accounting for the demagnetizing
effects, we find an almost 100\% superconducting shielding fraction,
suggestive of bulk SC, definitely confirmed by {\textmu}SR measurements (see below).

To determine the lower critical field $H_\mathrm{c1}$, the field-dependent
magnetization $M(H)$ of ThIrSi was measured at various temperatures up
to 6\,K. Figure~\ref{fig:chi}(b) shows the $M(H)$ curves at various
temperatures. The estimated $H_\mathrm{c1}$ values as a function of
temperature (accounting also for a demagnetization factor) are summarized
in Fig.~\ref{fig:chi}(c). The solid lines are fits to
$\mu_{0}H_\mathrm{c1}(T) =\mu_{0}H_\mathrm{c1}(0)[1-(T/T_{c})^2]$ and
yield a lower critical field $\mu_{0}H_\mathrm{c1}(0)$ = 6.8(1)\,mT
for ThIrSi.  

We also performed temperature-dependent magnetization measurements
$M(T,H)$ at various applied magnetic fields up to 7\,T. For each field,
$T_c$ was determined from the intersection of two straight lines drawn
on the normal and transition region. As shown in Fig.~\ref{fig:Hc2}(a),
upon increasing the magnetic field, the superconducting transition in
$M(T)$ becomes broader and shifts to lower temperatures.
Figure~\ref{fig:Hc2}(b) summarizes the superconducting transition
temperature $T_c$ vs.\ the applied magnetic field, 
as identified from the $M(H,T)$ data of ThIrSi. The $H_{c2}(T)$ was analyzed by means
of Ginzburg-Landau (GL)~\cite{Zhu2008}, Werthamer-Helfand-Hohenberg
(WHH)~\cite{Werthamer1966}, and two-band (TB) models~\cite{Gurevich2011}.
As shown in Fig.~\ref{fig:Hc2}, both the GL- and WHH models show large
deviations, leading to underestimated $H_\mathrm{c2}$ values at zero
temperature. Such discrepancy most likely hints at multiple superconducting
gaps in ThIrSi, as evidenced also by the positive curvature of $H_{c2}(T)$
at low fields, a typical feature of multigap superconductors, as e.g.,
MgB$_2$~\cite{Muller2001,Gurevich2004} or  Lu$_2$Fe$_3$Si$_5$~\cite{Nakajima2012}.
As shown in Fig.~\ref{fig:Hc2}(b), around $\mu_0H$ $\sim$2-3\,T, $H_{c2}(T)$
undergoes a clear change in curvature. The remarkable agreement of the
TB model with the experimental data across the full temperature range
is obvious 
%
\begin{figure}[!thp]
	\centering
	\includegraphics[width=0.49\textwidth,angle= 0]{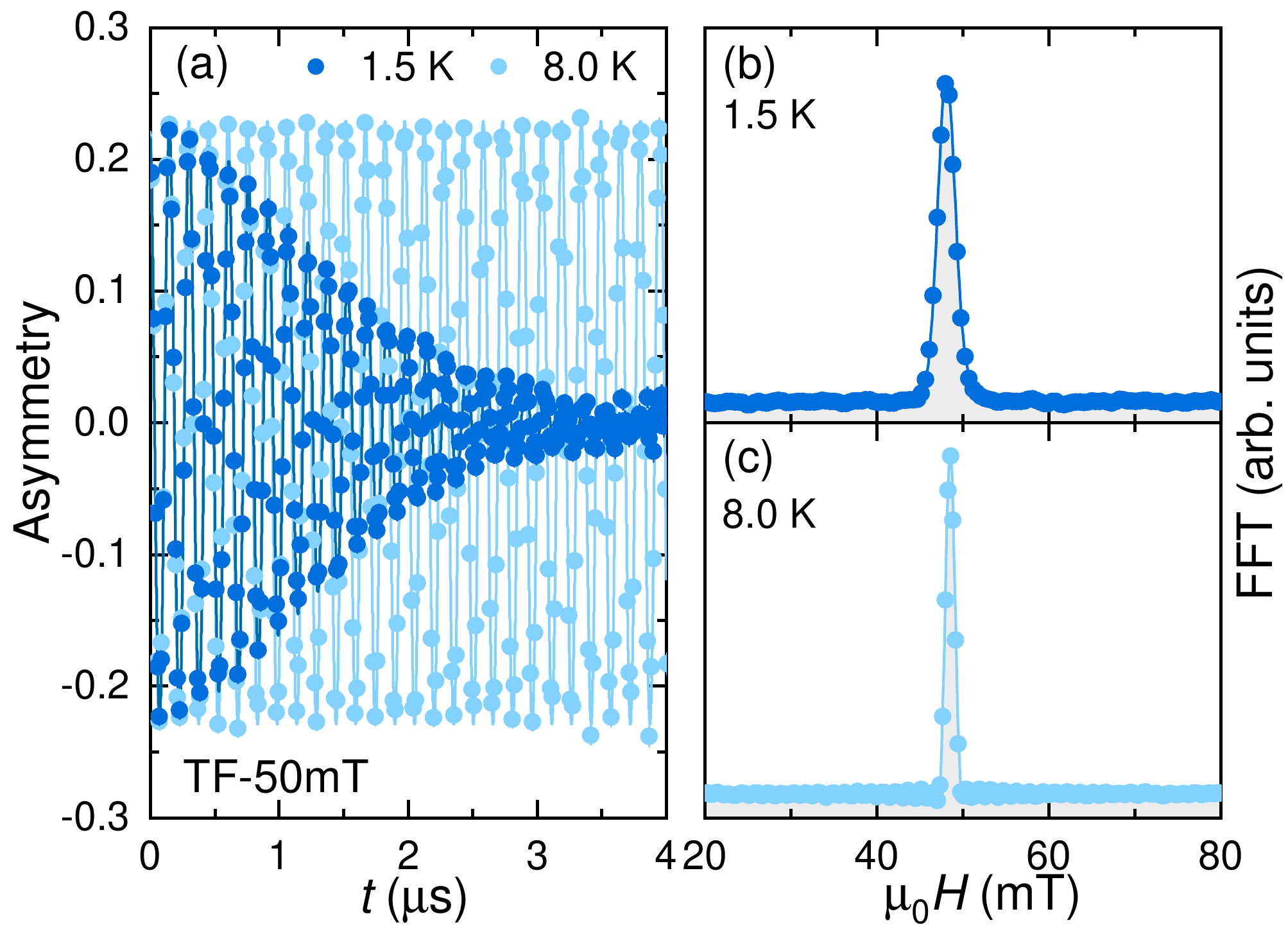}
	\caption{\label{fig:TF-muSR}(a) TF-{\textmu}SR spectra collected in an
	applied field of 50\,mT in both the superconducting- and normal
	states for ThIrSi. The real part of the fast Fourier transform of
	{\textmu}SR spectra is shown in (b) and (c) for 1.5\,K and 8.0\,K,
	respectively. Solid lines are fits to Eq.~\eqref{eq:TF_muSR}. The
	TF-200\,mT {\textmu}SR spectra shows similar features.}
\end{figure}
%
and it allows us to determine $\mu_0H_{c2}(0) = 10.8(2)$\,T and
$\xi(0) = 5.52(5)$\,nm. The lower critical field $\mu_{0}H_{c1}$ is
related to the magnetic penetration depth $\lambda$ and the coherence
length $\xi$ via $\mu_{0}H_{c1} = (\Phi_0 /4 \pi \lambda^2)[$ln$(\kappa)+ 0.5]$, 
where $\kappa$ = $\lambda$/$\xi$ is the GL parameter~\cite{Brandt2003}.
By using $\mu_{0}H_{c1} = 6.8(1)$\,mT and $\mu_{0}H_{c2} = 10.8(2)$\,T, 
the resulting magnetic penetration depth $\lambda_\mathrm{GL} = 333(3)$\,nm, 
is comparable to the experimental value 370(2)\,nm we determine from
TF-{\textmu}SR data (see below). The large GL parameter, $\kappa \sim 60$,
clearly indicates that ThIrSi is a type-II superconductor.
Lastly, when we compare $H_{c2}$ to the Pauli limit $H_\mathrm{P}$, given
by $\mu_{0}H_\mathrm{P} = \frac{\Delta_{0}}{\sqrt{2}\mu_\mathrm{B}} \approx 1.86\,\mathrm{[T/K]}\,T_{c}$,
we note that a $T_{c}$ of 6.4\,K corresponds to a $\mu_{0}H_\mathrm{P}$ of 11.9\,T.
Hence, the observed superconductivity in ThIrSi is well below the
Pauli limit, as expected of a conventional nodeless superconductor.

\subsection{{\textmu}SR study}
To investigate the superconducting pairing of ThIrSi, we carried out
systematic temperature-dependent TF-{\textmu}SR measurements in two magnetic
fields: 50 and 200\,mT. After cooling the sample in a transverse field,
the TF-{\textmu}SR spectra were collected upon heating. Representative datasets
for TF-50\,mT, taken in the superconducting- and  normal states of ThIrSi,
are shown in Fig.~\ref{fig:TF-muSR}, with the TF-200\,mT {\textmu}SR spectra
showing similar features. In the normal state, the {\textmu}SR asymmetry shows
essentially no damping, thus reflecting a uniform field distribution.
Conversely, in the superconducting state (here, at 1.5\,K), the
significantly enhanced damping reflects the inhomogeneous field
distribution due to the development of a flux-line lattice
(FLL)~\cite{Yaouanc2011,Amato1997,Blundell1999}. 
The broadening of the field distribution in the SC phase is clearly
visible in Fig.~\ref{fig:TF-muSR}(b) vs.\ Fig.~\ref{fig:TF-muSR}(c),
where the fast-Fourier-transform (FFT) spectra of the corresponding
TF-50\,mT {\textmu}SR data are shown. To properly describe the field
distribution, the time-dependent TF-{\textmu}SR asymmetry was modeled using:
\begin{equation}
	\label{eq:TF_muSR}
	A_\mathrm{TF}(t) = A_\mathrm{s} \cos(\gamma_{\mu} B_\mathrm{s} t + \phi) e^{- \sigma^2 t^2/2} +
	A_\mathrm{bg} \cos(\gamma_{\mu} B_\mathrm{bg} t + \phi).
\end{equation}
Here $A_\mathrm{s}$ (98\%), $A_\mathrm{bg}$ (2\%) and $B_\mathrm{s}$,
$B_\mathrm{bg}$ are the initial asymmetries and local fields sensed by
implanted muons in the sample and sample holder, $\gamma_{\mu}/2\pi = 135.53$\,MHz/T 
is the muon gyromagnetic ratio, $\phi$ is a shared initial phase, and
$\sigma$ is a Gaussian relaxation rate reflecting the field distribution
inside the sample. 
%
\begin{figure}[!thp]
	\centering
	\includegraphics[width=0.46\textwidth,angle= 0]{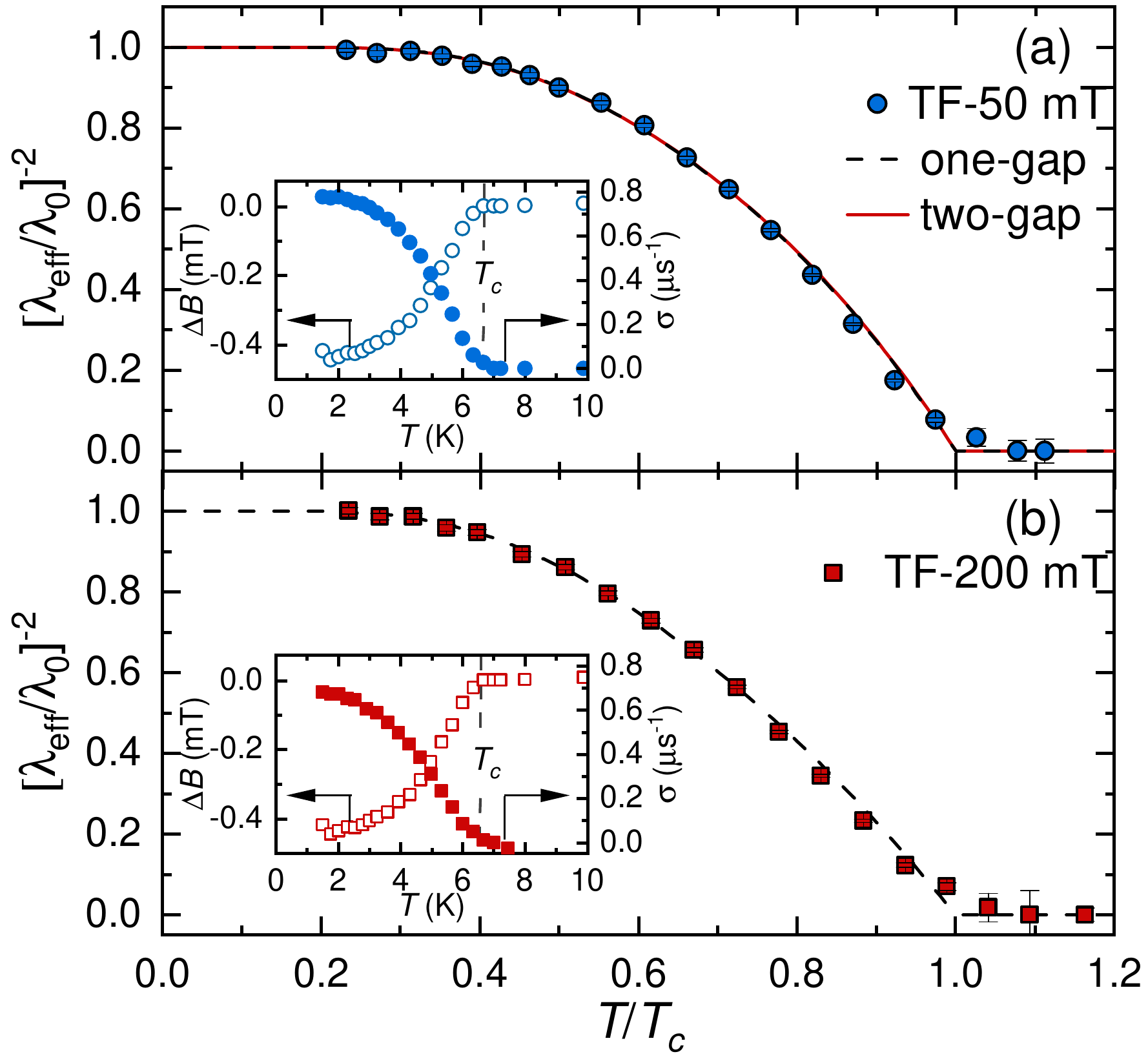}
	\caption{\label{fig:superfluid}Superfluid density vs temperature, as
	determined from	TF-{\textmu}SR measurements in an applied magnetic field
	of 50\,mT (a) and 200\,mT (b). The insets show the diamagnetic shift
	$\Delta$$B(T) $(left-axis) and muon-spin relaxation rates $\sigma(T)$
	(right-axis). Here, $\Delta B = B_\mathrm{s} - B_\mathrm{appl.}$,
	where $B_\mathrm{appl.} = 50$ or 200\,mT. 
	The dashed- and solid lines represent fits to a fully-gapped
	model with a single gap and two gaps, respectively. 
	The fit parameters are listed in Table~\ref{tab:gapvalues}.}
\end{figure}
%
The derived $\sigma$ are small and temperature-independent in the
normal state, but below $T_c$ they start to increase due to the onset
of FLL and the increased superfluid density. Simultaneously, a
diamagnetic shift $\Delta$$B$ appears below $T_c$ (see inset in
Fig.~\ref{fig:superfluid}).
In the superconducting state, $\sigma$ includes contributions from both 
the FLL ($\sigma_\mathrm{sc}$) and a smaller, temperature-independent
relaxation, due to the nuclear moments ($\sigma_\mathrm{n}$). 
Considering the constant nuclear relaxation rate in the narrow
temperature range investigated here, confirmed also by ZF-{\textmu}SR
measurements (see below), the superconducting Gaussian relaxation rate
can be extracted using 
$\sigma_\mathrm{sc} = \sqrt{\sigma_\mathrm{eff}^{2} - \sigma^{2}_\mathrm{n}}$.

Since $\sigma_\mathrm{sc}$ is directly related to the effective magnetic 
penetration depth and, thus, to the superfluid density, the
superconducting gap and its symmetry can be investigated by measuring
the temperature-dependent $\sigma_\mathrm{sc}$. Then, the effective
magnetic penetration depth $\lambda_\mathrm{eff}$ can be obtained by using 
$\sigma_\mathrm{sc}^2(T)/\gamma^2_{\mu} = 0.00371\Phi_0^2/\lambda_\mathrm{eff}^4(T)$~\cite{Barford1988,Brandt2003}.

The normalized inverse-square of the effective magnetic penetration
depth [proportional to the superfluid density, i.e.,
$\lambda_\mathrm{eff}^{-2}(T) \propto \rho_\mathrm{sc}(T)$] vs.\ the
reduced temperature $T/T_c$ is presented in Fig.~\ref{fig:superfluid}(a)
and \ref{fig:superfluid}(b) for TF-50\,mT and TF-200\,mT {\textmu}SR spectra,
respectively. In both cases, the superfluid densities are temperature
invariant below 1/3$T_c$, thus indicating the absence of low-energy
excitations and, hence, a fully-gapped superconducting state in ThIrSi.
Such nodeless SC is also confirmed by NMR measurements (see below).
Consequently, the superfluid density $\rho_\mathrm{sc}(T)$ was analyzed
by means of a fully-gapped $s$-wave model:
\begin{equation}
	\label{eq:rhos}
	\rho_\mathrm{sc}(T) = \frac{\lambda_\mathrm{eff}^{-2}(T)}{\lambda_0^{-2}} = 1 + 2\int^{\infty}_{\Delta(T)} \frac{\partial f}{\partial E} \frac{EdE}{\sqrt{E^2-\Delta^2(T)}}.
\end{equation}
Here, $f = (1+e^{E/k_\mathrm{B}T})^{-1}$ is the Fermi function;
$\Delta(T)$ is the  su\-per\-con\-duc\-ting\--gap function, assumed to
follow $\Delta(T) = \Delta_0 \mathrm{tanh} \{1.82[1.018(T_\mathrm{c}/T-1)]^{0.51} \}$~\cite{Tinkham1996,Carrington2003}; $\lambda_0$ and $\Delta_0$ are the magnetic
penetration depth and the superconducting gap at 0\,K, respectively.  
As shown by the solid lines in Fig.~\ref{fig:superfluid}, the $s$-wave 
model describes $\rho_\mathrm{sc}(T)$ very well across the entire
temperature range with the fit parameters:$\Delta_0$ = 2.10(5) and
1.90(5)\,$k_\mathrm{B}T_c$, and $\lambda_0$ = 370(2) and 397(2)\,nm for
TF-50\,mT and TF-200\,mT {\textmu}SR spectra, respectively.

%
\begin{figure}[!thp]
	\centering
	\includegraphics[width=0.48\textwidth,angle= 0]{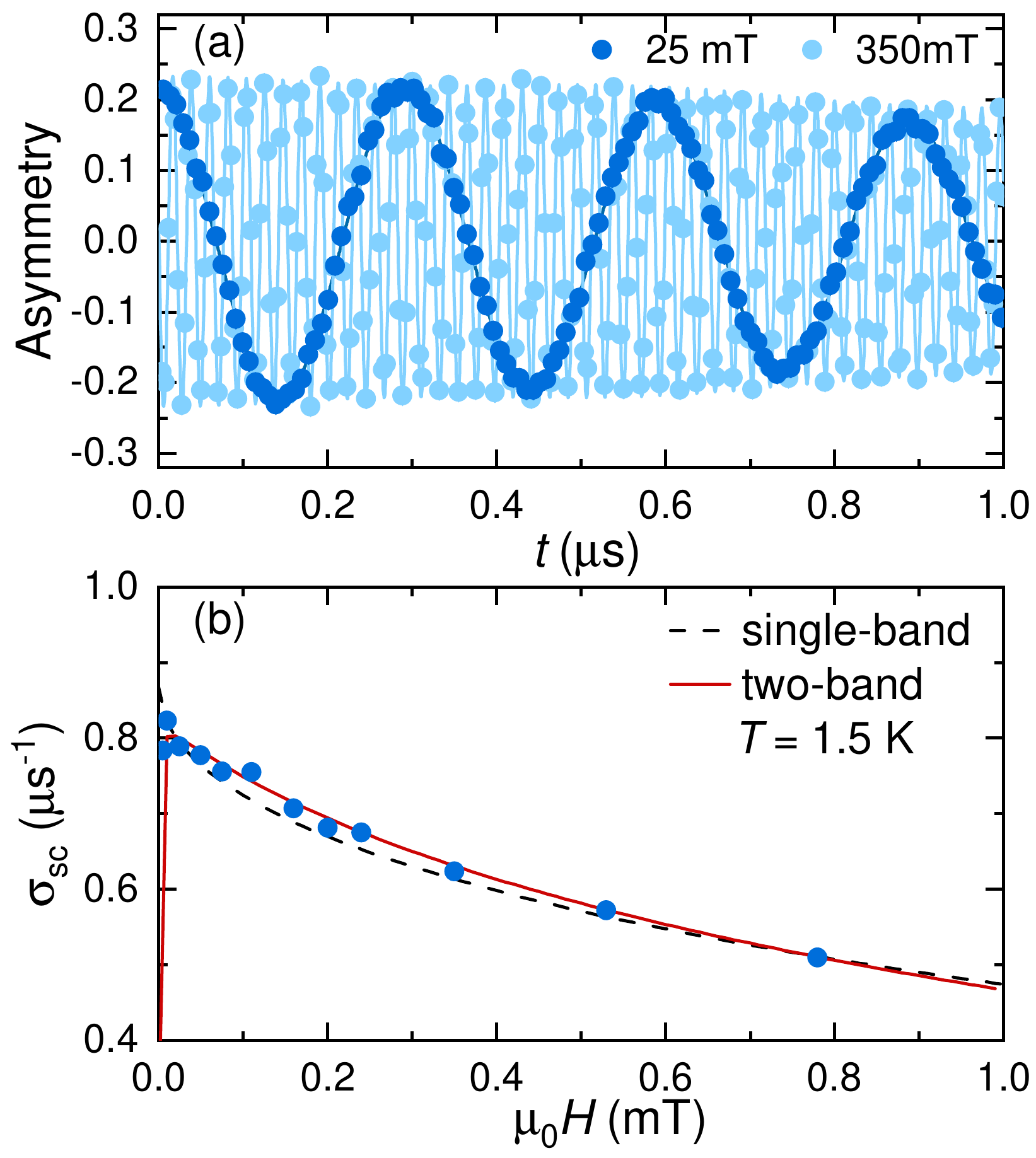}
	\caption{\label{fig:lambda2}(a) TF-{\textmu}SR time spectra measured in
	the superconducting state of ThIrSi (at $T = 1.5$\,K) in a field of
	25 and 350\,mT. (b) Field-dependent superconducting Gaussian relaxation
	rate $\sigma_\mathrm{sc}(H)$. Dashed- and solid lines represent fits
	to the single- and two-band models, respectively. The goodness-of-fit
	values are $\chi^2_\mathrm{r}$ = 3.2 (two-band model) and 8.4
	(single-band model). In both cases, data below $H_\mathrm{c1}$ were
	excluded when evaluating $\chi^2_\mathrm{r}$.}
\end{figure}

Since the $H_\mathrm{c2}(T)$ data (see Fig.~\ref{fig:Hc2}) show some
features of multigap SC, we analyzed the superfluid density also with
the so-called $\alpha$-model. In this case, the superfluid density can
be described by $\rho_\mathrm{sc}(T) = \alpha \rho_\mathrm{sc}^{\Delta^\alpha}(T) + (1-\alpha) \rho_\mathrm{sc}^{\Delta^\beta}(T)$, where $\rho_\mathrm{sc}^{\Delta^\alpha}$
and $\rho_\mathrm{sc}^{\Delta^\beta}$ are the superfluid densities
related to the first ($\Delta^\alpha$) and second ($\Delta^\beta$) gap,
and $\alpha$ is a relative weight. For each gap, $\rho_\mathrm{sc}(T)$
is given by Eq.~\eqref{eq:rhos}. As shown by the solid line in
Fig.~\ref{fig:superfluid}(a), by using the same $\lambda_0$ value and
by fixing $w$ to 0.3 (as derived from the field-dependent TF-{\textmu}SR
measurements reported below), the resulting gap values are
$\Delta^\alpha = 1.90(5)$\,$k_\mathrm{B}$$T_c$ and
$\Delta^\beta =  2.20(5)$\,$k_\mathrm{B}$$T_c$. 
However, as shown in Fig.~\ref{fig:superfluid}, the two-gap fit
(solid line) is virtually indistinguishable from the one-gap fit
(dashed line).

This circumstance might be attributed to the relatively small weight
of the second gap, as well as to the comparable gap-energy sizes, both
factors which make it difficult to discriminate between a single- and
a two-gap superconductor based on the temperature-dependent superfluid
density alone. In this case, measurements of the field-dependent
superconducting Gaussian relaxation rate $\sigma_\mathrm{sc}(H)$
provides suitable alternatives to disentangle the two cases, since
the respective datasets are expected to show distinct field responses 
in a two-gap vs.\ a single-gap superconductor~\cite{Shang2020MoPB,Shang2022NbReSi}. 
Based on this, to ascertain the possibility of a multigap SC in ThIrSi,
we performed TF-{\textmu}SR measurements at base temperature (1.5\,K) at
different applied fields, up to 780\,mT. Figure~\ref{fig:lambda2}(a)
shows two representative TF-{\textmu}SR datasets, collected at 25 and 350\,mT,
with the spectra in other applied fields showing similar features. 
Also in this case, we fitted the data by means of Eq.~\eqref{eq:TF_muSR},
with the resulting superconducting Gaussian relaxation rates
$\sigma_\mathrm{sc}$ vs the applied magnetic field being summarized in
Fig.~\ref{fig:lambda2}(b). $\sigma_\mathrm{sc}(H)$ was analyzed using
both a single- and a two-band model. In the latter case, each band is
characterized by its own superconducting coherence length [i.e., $\xi_1(0)$
and $\xi_2(0)$], while a weight $w$ accounts for the contribution of
the first band [$\xi_1(0)$] to the total superfluid density, akin to the
two-gap model in Fig.~\ref{fig:superfluid}~\cite{Serventi2004,Khasanov2014}.
As shown in Fig.~\ref{fig:lambda2}(b), the two-band model (solid line)
shows a better agreement with the data, here reflected in a smaller
$\chi^2_\mathrm{r}$ value, and it yields these best-fit parameters:
$w = 0.3$, $\xi_1(0) = 12.0$\,nm, $\xi_2(0) = 5.4$\,nm, and
$\lambda_0 = 357$\,nm.
The derived $\lambda_0$ is consistent with that obtained from the
$\rho_{sc}(T)$ analysis in Fig.~\ref{fig:superfluid} (370\,nm).{}
Finally, the upper critical field of 11.2\,T, calculated from the
coherence length of the second band $\xi_2(0)$, is comparable with the
$H_\mathrm{c2}$ value determined from the $M(T,H)$ data (see
Fig.~\ref{fig:Hc2}).
As for the first band, the critical field of 2.3\,T, calculated from
$\xi_1(0)$, is in good agreement with the field value where
$H_\mathrm{c2}(T)$ changes its slope [see Fig.~\ref{fig:Hc2}(b)].
Note that, in case of the NMR measurements, here performed in a magnetic
field of 3\,T, the required high field can suppress the smaller gap.
As a consequence, the material is expected to appear more like a
single-gap superconductor.

%
\begin{figure}[!tp]
	\centering
    \vspace{-1ex}
	\includegraphics[width=0.45\textwidth]{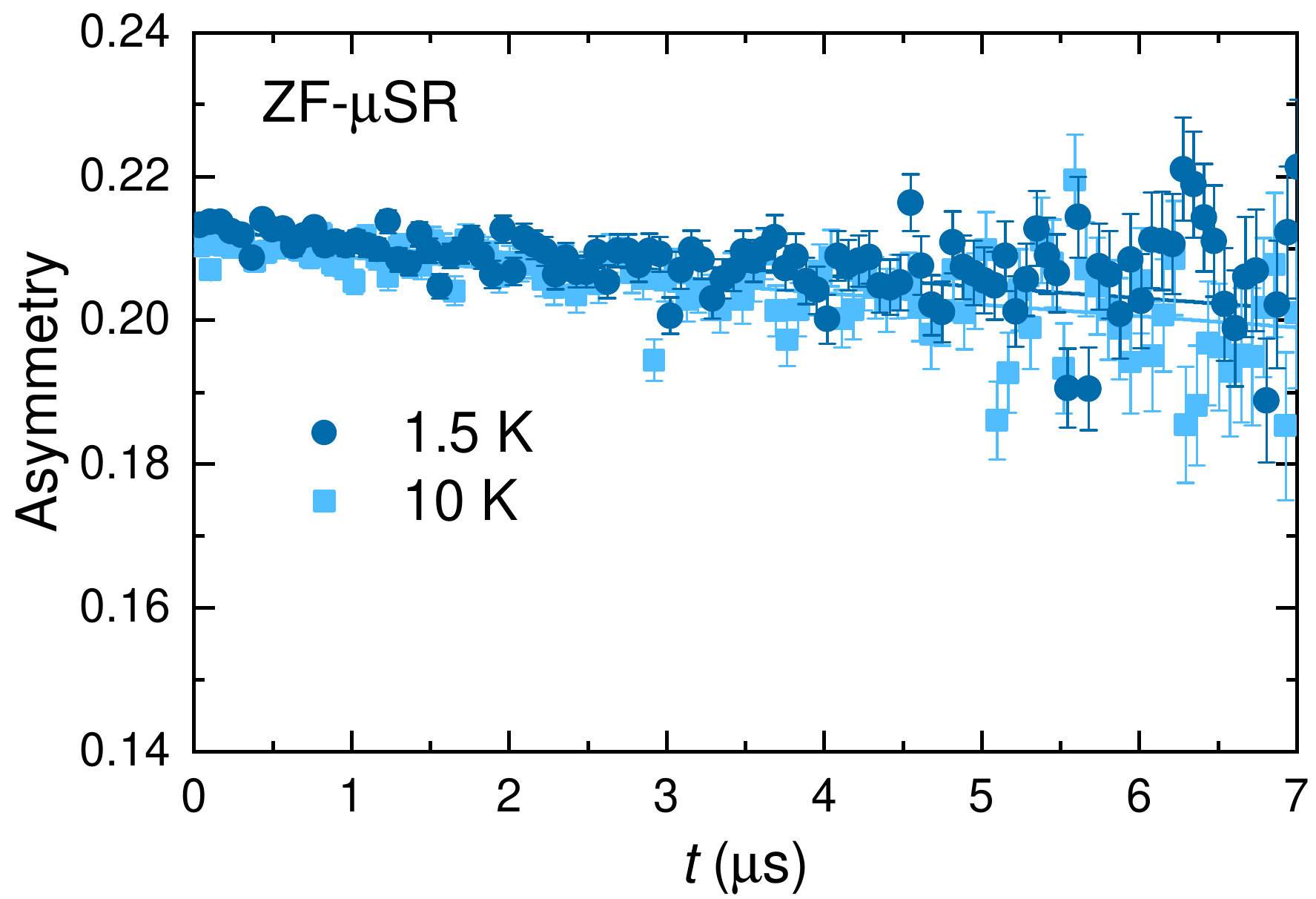}
	\caption{\label{fig:ZF-muSR}ZF-{\textmu}SR spectra collected in the
	superconducting- (1.5\,K) and the normal (10\,K) states of ThIrSi.
	The practically overlapping datasets indicate the absence of TRS
	breaking, whose occurrence would have resulted in a stronger decay
	in the 0.3-K case.}
\end{figure}
%
We also performed zero-field (ZF-) {\textmu}SR measurements in the normal-
and superconducting states of ThIrSi, in order to reveal a possible
breaking of the time-reversal symmetry (TRS), in turn implying an
unconventional SC. As shown in Fig.~\ref{fig:ZF-muSR}, neither coherent
oscillations nor fast decays could be identified in the spectra collected
below- (1.5\,K) and above $T_c$ (10\,K), thus excluding any type of
magnetic order or fluctuations.
In nonmagnetic materials, in the absence of applied fields, the
depolarization of muon spins is mainly determined by the randomly
oriented nuclear magnetic moments. In ThIrSi, the depolarization shown
in Fig.~\ref{fig:ZF-muSR} is more consistent with a Lorentzian decay.
This suggests that the internal fields sensed by the implanted muons
arise from the diluted (and tiny) nuclear moments present in ThIrSi. 
Thus, the solid lines in Fig.~\ref{fig:ZF-muSR} are fits to a
Lorentzian Kubo-Toyabe relaxation function 
$A(t) = A_\mathrm{s}[\frac{1}{3} + \frac{2}{3}(1 - \Lambda_\mathrm{ZF} t) \mathrm{e}^{- \Lambda_\mathrm{ZF} t}] + A_\mathrm{bg}$. 
Here, $A_\mathrm{s}$ and $A_\mathrm{bg}$ are the same as in the 
TF-{\textmu}SR case [see Eq.~\eqref{eq:TF_muSR}], while $\Lambda_\mathrm{ZF}$
represents the ZF Lorentzian relaxation rate. The derived relaxation
rates in the normal- and the superconducting state are almost identical,
i.e., $\Lambda_\mathrm{ZF}$  = 0.0057(7)\,{\textmu}s$^{-1}$ at 1.5\,K and
$\Lambda_\mathrm{ZF}$ = 0.0060(8)\,{\textmu}s$^{-1}$ at 10\,K, here reflected
in overlapping datasets. The lack of an additional {\textmu}SR relaxation
below $T_c$ excludes a possible TRS breaking in the superconducting
state of ThIrSi. As we show below, the conventional nature of SC in
ThIrSi is further supported by the exponential dependence of the NMR
relaxation rate and by a clear drop in the NMR shift below $T_c$.

%
\begin{figure}[t]
	\centering
	\includegraphics[width=0.48\textwidth,angle=0]{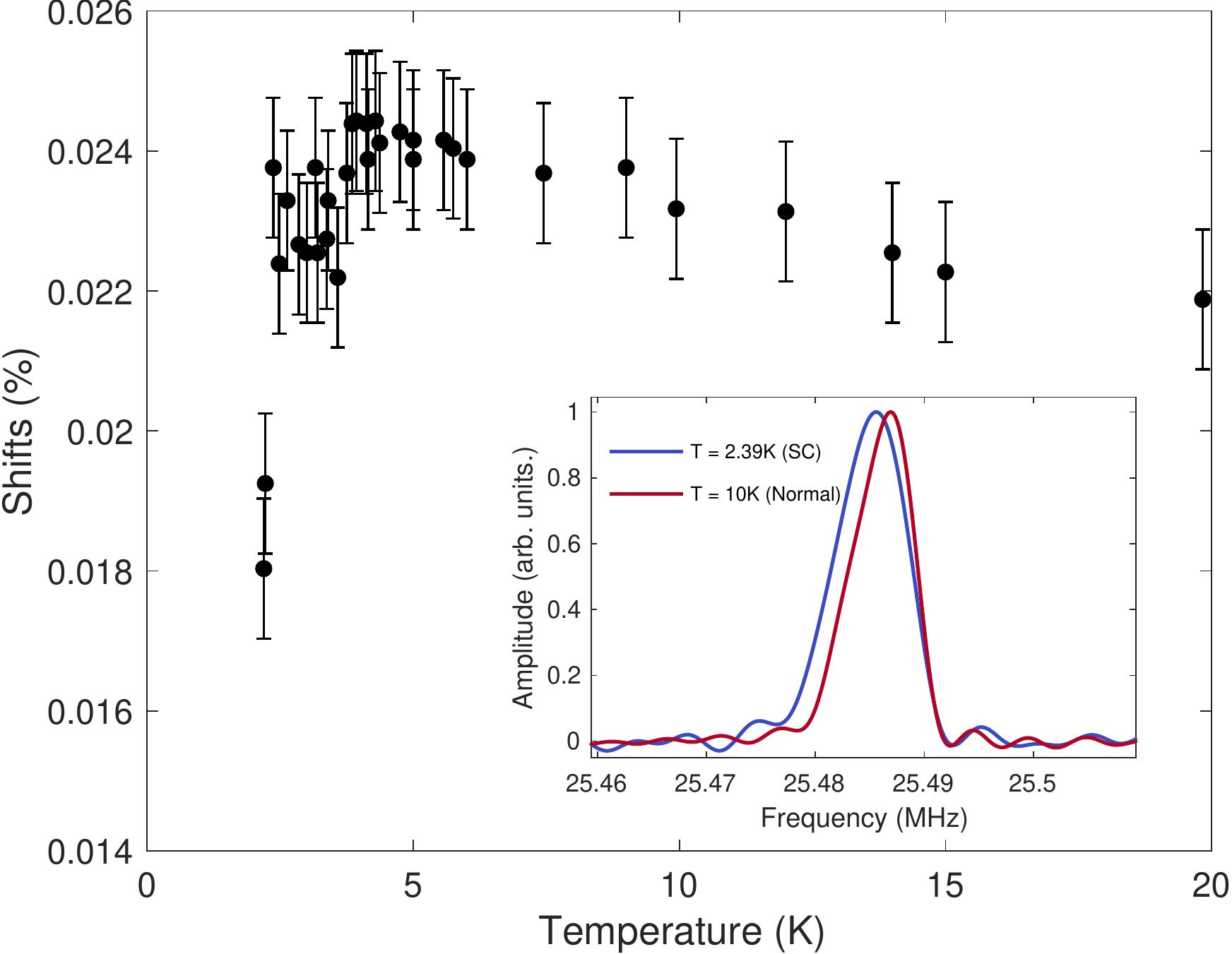}
	\vspace{-2ex}%
	\caption{\label{fig:lines_and_shifts}Evolution of the $^{29}$Si NMR
	shift with temperature. As typical for $s$-wave superconductors,
	the shift decreases below $T_c$. Inset: Representative $^{29}$Si NMR
    line shapes in the superconducting- and the normal state, collected
    in a magnetic field of 3\,T.}
\end{figure}
%

\subsection{$^{29}$Si NMR study}
\begin{figure}[t]
	\centering
	\includegraphics[width=0.48\textwidth,angle=0]{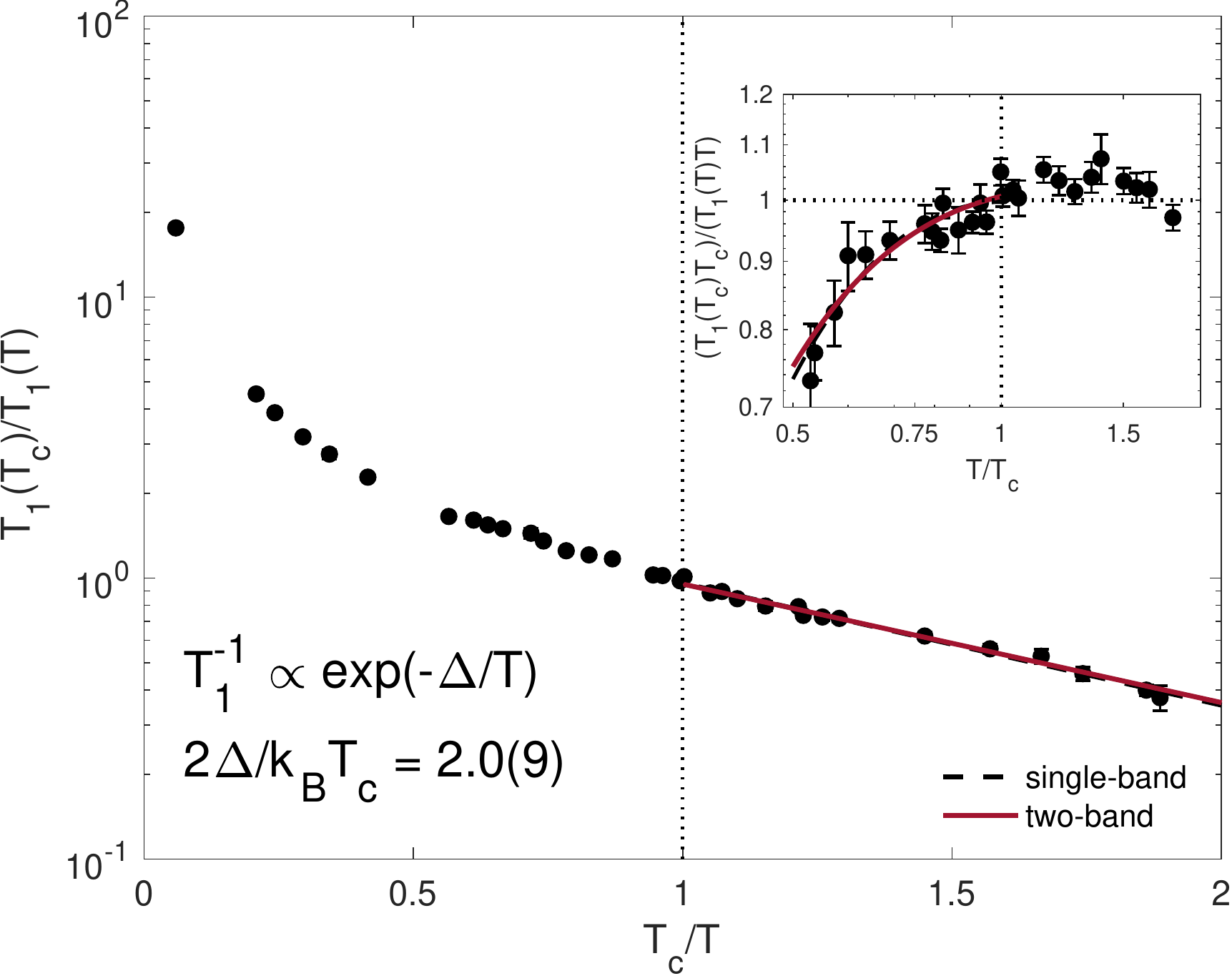}
	\vspace{-2ex}%
\caption{\label{fig:relaxation}Normalized NMR relaxation rate vs the
scaled temperature (measured at $\mu_0H = 3$\,T). The decay follows an
exponential law, $T_1^{-1} \propto \exp(-\Delta/k_\mathrm{B}T)$, typical
of $s$-wave superconductors with a fixed gap $\Delta$.
Inset: Scaled Korringa product vs the scaled temperature. In both cases,
the two-band fit (solid line) is hardly distinguishable from the
single-band fit (dashed line), most likely reflecting the suppression
of the smaller gap by the magnetic field.}
\end{figure}

From the basic theory of NMR in superconducting materials (see, e.g.,
Ref.~\onlinecite{MacLaughlin1976, *Walstedt2008}), it is known that
conventional BCS superconductors exhibit three key signatures below
$T_{c}$:\\[2mm]
\noindent
\textit{a.} A reduced Knight shift $K$ with respect to the normal-state value.\\
\textit{b.} An exponential decrease of the relaxation rate $T_1^{-1}(T)$.\\
\textit{c.} The appearance of a Hebel-Slichter (HS) coherence peak in the
Korringa product $(T_{1}T)^{-1}$ just below $T_{c}$.\\[2mm]
In ThIrSi, the decreasing Knight shift below $T_{c}$ is clearly evident
in Fig.~\ref{fig:lines_and_shifts}, while the exponential decrease of
$T^{-1}(T)$ below $T_{c}$ is made obvious by the semilogarithmic scale
in Fig.~\ref{fig:relaxation}. The only missing signature is the
Hebel-Slichter coherence peak, which is conspicuously absent in
Fig.~\ref{fig:relaxation} (see inset). However, we recall that the
Hebel-Slichter peak is suppressed also in other conventional
noncentrosymmetric superconductors, such as NbReSi~\cite{Shang2022NbReSi}
and W$_3$Al$_2$C~\cite{Tay2022}, which also are fully gapped.
A comparison of $1/T_{1}T$ vs $T$ in different NCSCs is given in our
previous paper ~\cite{Tay2022}. It appears that, experimentally the
Hebel-Slichter peak is either completely absent or strongly suppressed
in the NCSCs studied. Hence, it is not surprising that we do not observe
this feature in ThIrSi either. 
Furthermore, the two main signatures of nodal (triplet)
superconductivity~\cite{Bauer2012}, namely a power-law dependence of
relaxation rates and a temperature-independent Knight shift below
$T_c$, are clearly missing in our case. Hence, in spite of the lack of
an HS peak, nodeless superconductivity remains the most convincing
scenario compatible with our NMR data.

%

Next, we move on to discuss the details of the gap structure
found via NMR. As can be seen in Fig.~\ref{fig:relaxation}, the NMR
relaxation rate follows a single-exponential law, here depicted by the
dashed line. However, the gap resulting from the single-exponential fit
is $2\Delta/k_\mathrm{B}T_{c} = 2.0(9)$, which is significantly lower
than that predicted by the BCS theory, i.e., $2\Delta/k_\mathrm{B}T_{c} = 3.5$.
We attempted to fit the data using a two-exponential function, but the
second exponential does not change the result. The value of the main gap
is still $2\Delta_{a}/k_\mathrm{B}T_{c} = 2.0(9)$ (within experimental
error), while the second exponential has a negligibly small weight.
Although apparently NMR seems to exclude the possibility of a two-gap
superconductivity, it is quite likely that the magnetic field required
for the NMR experiments might suppress the smaller gap.

\begin{table}
    \centering
    \caption{\label{tab:gapvalues} Superconducting gap values of ThIrSi
   	compared to those of Y$_2$C$_3$ and La$_2$C$_3$, as obtained from
   	NMR and {\textmu}SR measurements. The NMR $T_{c}$ and $\Delta$
   	values refer to the 3-T magnetic field.
   	The experimental uncertainties are of
   	the order of the last digit.}
    \begin{tabular}{p{18mm}p{18mm} c c c}
     \toprule 
     Compound & Technique & $2\Delta_{\alpha}/k_\mathrm{B}T_c$  &$2\Delta_{\beta}/k_\mathrm{B}T_c$   &$\alpha$/$\beta$\\
     \midrule
      ThIrSi & {\textmu}SR & 4.4 &3.8  &0.3/0.7 \\
      ThIrSi &  NMR & 2.0 & NA    & NA\\
      Y$_2$C$_3$~\cite{Kuroiwa2008}  & {\textmu}SR & 4.9 & 1.1  &0.86/0.14\\ 
      Y$_2$C$_3$~\cite{Harada2007}   & NMR     & 5.0 & 2.0  & 0.75/0.25\\
      La$_2$C$_3$~\cite{Kuroiwa2008} & {\textmu}SR & 5.6 & 1.3  &0.38/0.62 \\
      \bottomrule
     \end{tabular}
\end{table}

\subsection{Discussion}
In ThIrSi, we find that the datasets resulting from three different
techniques: i.e., magnetometry, {\textmu}SR, and NMR, agree in
indicating a fully-gapped singlet superconducting state. However, the
situation is less clear with respect to the characteristics of the
$s$-wave type pairing, i.e., whether it involves quasiparticles in a
one- or two-band configuration.

The temperature dependence of the upper critical field $H_{c2}(T)$, 
obtained from $M(T,H)$ measurements described above and shown in
Fig.~\ref{fig:Hc2}(b), indicates a much better fit when employing a 
two-band model than when using the classical models (that are based on
assuming a single-band of quasiparticles). 
The NMR results suggest that a single-band model fits the
corresponding data adequately. However, the field-dependent {\textmu}SR
results, are best fit using a two-band $(s+s)$-wave gap configuration.
Although at first sight the NMR results might seem contradictory, we
recall that NMR typically requires high magnetic fields (3\,T, in our
case), which may easily suppress the smaller superconducting gap and,
thus, lead to an apparent single-band scenario.

Also recent DFT calculations~\cite{ptok2019electronic} seem to support
the two-band interpretation since, in the ThIrSi case, they indicate:
(i) a mix of isotropic and anisotropic bands near the Fermi level, a
requirement for two-band superconductivity, and (ii) large band
splittings resulting from spin-orbit coupling (SOC). The combination
of multiple bands and large SOC is fully compatible with the two-band
model, while it renders the simple single-gap $s$-wave model unlikely.

However, also the two-gap model runs into theoretical problems. In the
two-gap picture, we observe that the SC gap values obtained for ThIrSi
are relatively close to each other and the ratio of the relative weights
is close to 0.5/0.5.
These fit parameters are substantially different from those regarding
the sesquicarbides (see Table~\ref{tab:gapvalues}), the latter appearing
as more clearcut examples of two-gap superconductors. Indeed, contrary
to ThIrSi, in the sesquicarbides both NMR and {\textmu}SR data exhibit
two-band features. In this sense, it is a still open issue whether
ThIrSi is a two-band superconductor.

In general, if the weight of the second gap is relatively small and the
gap sizes are not significantly different, this makes it difficult to
discriminate between a single- and a two-gap superconductor based on
temperature-dependent superconducting properties.
For ThIrSi, the weight of the second gap $w = 0.3$ and the gap sizes
are quite similar (see Table~\ref{tab:gapvalues}). As a consequence,
the multigap feature is less evident in the temperature-dependent
superfluid density (see Fig.~\ref{fig:superfluid}). From the analysis
of $H_{c2}(T)$ using a two-band model, the derived inter-band and
intra-band couplings are $\lambda_{12} = 0.03$ and
$\lambda_{11} \sim \lambda_{22} = 0.25$ for ThIrSi. Such an inter-band
coupling is much smaller than the intra-band coupling, a circumstance
which makes the gaps to open at different electronic bands, less
distinguishable compared with other multiband superconductors (see
Table~\ref{tab:gapvalues})~\cite{Kogan2009}. However, the underlying
multigap SC feature of ThIrSi is reflected in its upper critical fields
$H_\mathrm{c2}(T)$ (see Fig.~\ref{fig:Hc2}). The measurement of the
field-dependent superconducting Gaussian relaxation rate
$\sigma_\mathrm{sc}(H)$ also provides evidence of multiband SC (see
Fig.~\ref{fig:lambda2}, which shows a distinct field response compared
to a single-gap superconductor~\cite{Shang2019c,Shang2020MoPB}).

While the discrepant SC-gap values remain an open problem,
the lack of TRS breaking deduced from ZF-{\textmu}SR, excludes the
possibility of spin-triplet- or other non $s$-wave superconductivity
mechanisms in ThIrSi.

\section{Conclusion}\enlargethispage{8pt}
\label{ssec:Sum}
In summary, we presented the results of an extensive study of the
properties of the non-centrosymmetric superconductor ThIrSi, by
employing magnetometry, as well as NMR and {\textmu}SR techniques. 
Experimental data confirm the formation of a nodeless gap configuration
in the superconducting state. Nevertheless, data interpretation does
not allow for a clear decision whether ThIrSi adopts a single- $s$
or a two-band $(s+s)$ superconducting state.
Future experimental and theoretical investigations of similar materials
may help to clarify the link between the non-centrosymmetry of the
crystal structure and multiband superconductivity. 
In this respect, DFT calculations show that ThNiSi and ThPtSi have
multiple bands near (or crossing) the Fermi
level~\cite{ptok2019electronic} and, hence, are suitable candidate
materials to search for multigap superconductivity, once they can be
shown to adopt a superconducting state at low temperatures. 

\vspace{1pt}
\begin{acknowledgments}
This work was supported from the Natural Science Foundation of 
Shanghai (Grants No.\ 21ZR1420500 and 21JC1402300), 
This work was supported by the Natural Science Foundation of Shanghai 
(Grants No.\ 21ZR1420500 and 21JC140\-2300), the Natural Science
Foundation of Chongqing (Grant No.\ CSTB2022NSCQ-MSX1678), the Fundamental
Research Funds for the Central Universities, and the Schweizerische 
Nationalfonds zur F\"{o}rderung der Wis\-sen\-schaft\-lichen For\-schung 
(SNF) (Grants No.\ 200021\_188706 and 206021\_139082). 
Work at Los Alamos was supported by the U.S.\ Department of Energy,
Office of Science, National Quantum Information Science Research Centers,
Quantum Science Center. F.B.S.\ was supported by FAPESP under Grants
No.\ 2016/11565-7 and 2018/20546-1. 
We acknowledge the allocation of beam time at the Swiss muon source 
(GPS {\textmu}SR spectrometer).
\end{acknowledgments}

\bibliography{ThIrSi.bib}

\end{document}